\documentclass{article}
\pdfoutput=1

\usepackage[preprint,  nonatbib]{neurips_2023}

\usepackage[utf8]{inputenc} 
\usepackage[T1]{fontenc}    
\usepackage{hyperref}       
\usepackage{url}            
\usepackage{booktabs}       
\usepackage{amsfonts}       
\usepackage{amsmath}
\usepackage{mathtools}
\usepackage{nicefrac}       
\usepackage{microtype}      
\usepackage{xcolor}         
\usepackage{tabularx}
\usepackage{wrapfig}
\usepackage{enumitem}
\usepackage{lipsum,booktabs}
\usepackage[pdftex]{graphicx}
\usepackage{verbatim}

\usepackage{bmpsize}

\DeclareMathOperator*{\argmax}{arg\,max}

\usepackage{xcolor}
\newcommand{\todo}[1]{{\color{red}[TODO: #1]}}

\newcommand{\dset}{{{MultiVENT}}}
\newcommand{\dsetsp}{{{MultiVENT} }}

\usepackage[capitalize]{cleveref}
\crefname{section}{Sec.}{Secs.}
\Crefname{section}{Section}{Sections}
\Crefname{table}{Table}{Tables}
\crefname{table}{Tab.}{Tabs.}
\Crefname{figure}{Figure}{Figures}
\crefname{figure}{Fig.}{Figs.}

\makeatother

\makeatletter
\newcommand{\thickhline}{%
    \noalign {\ifnum 0=`}\fi \hrule height 1pt
    \futurelet \reserved@a \@xhline
}
\newcolumntype{"}{@{\hskip\tabcolsep\vrule width 1pt\hskip\tabcolsep}}
\makeatother

\title{MultiVENT: \protect\\ Multilingual Videos of Events \protect\\ with Aligned Natural Text}

\author{Kate Sanders*\hspace{7mm}David Etter*\hspace{7mm}Reno Kriz*\hspace{7mm}Benjamin Van Durme\\
Johns Hopkins University\\
Human Language Technology Center of Excellence\\
\texttt{\{ksande25, detter2, rkriz1, vandurme\}@jhu.edu}\\
}

\begin{document}

\maketitle
\def\thefootnote{*}\footnotetext{Equal contribution.}\def\thefootnote{\arabic{footnote}}

\begin{abstract}\label{abstract}

Everyday news coverage has shifted from traditional broadcasts towards a wide range of presentation formats such as first-hand, unedited video footage. Datasets that reflect the diverse array of multimodal, multilingual news sources available online could be used to teach models to benefit from this shift, but existing news video datasets focus on traditional news broadcasts produced for English-speaking audiences. We address this limitation by constructing \dset, a dataset of multilingual, event-centric videos grounded in text documents across five target languages. \dsetsp includes both news broadcast videos and non-professional event footage, which we use to analyze the state of online news videos and how they can be leveraged to build robust, factually accurate models. Finally, we provide a  model for complex, multilingual video retrieval to serve as a baseline for information retrieval using \dset.
\end{abstract}

\section{Introduction}\label{introduction}

Information dissemination for current events has traditionally consisted of professionally collected and produced materials, leading to large collections of well-written news articles and high-quality videos. As a result, such materials form the basis for significant prior work in content analysis and retrieval \cite{yang2003videoqa, jahagirdar2023watching, araujo2015stanford, ellis2014we, wu2023newsnet}. Meanwhile, a high volume of event-centric content today is generated by non-professionals, such as on-the-scene witnesses to events who hastily capture videos and upload them to the internet without further editing. We propose that this contemporary landscape of news content can be leveraged by models to produce a more comprehensive understanding of events. News agencies have adapted to this shift, often collecting and incorporating this online content into official broadcasts, but news video datasets still do not typically address this new domain of event coverage. 

In addition to focusing on traditional news sources, existing news video datasets predominantly consider content produced in English. This is consistent with common practices in video dataset collection: Collected videos and captions are recorded in English, and when multilinguality is considered, it is achieved by directly translating captions and transcripts \cite{wang2019vatex, lei2021mtvr, rouditchenko2023c2kd, huang2021multilingual}. Because this data is originally produced for English speaking audiences, these multilingual datasets can contain unwanted content biases like "translationese" \cite{baroni2006new, koppel2011translationese}. As event-centric video content produced in other languages makes up a large portion of news videos online, we argue that including organic, multilingual content is necessary for a diverse and perspective-agnostic sampling of event coverage. 

\begin{figure}
 \includegraphics[width=\textwidth]{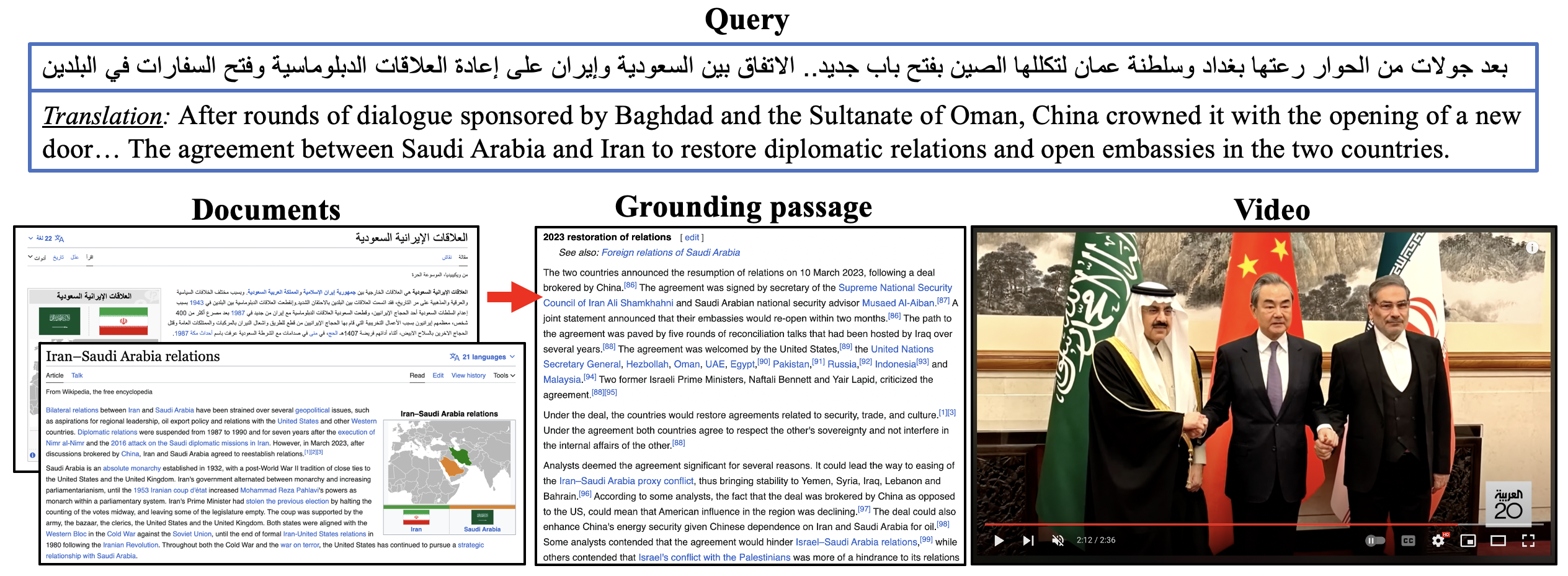}
  \caption{A sample video-text pair from \dset. Every event-centric video is paired with a corresponding video description and a long-form text document describing the event, both in the same language as the video. If the language is not English, the video is also paired with a corresponding English document.}
 \label{fig:title}
\end{figure}

With these ideas in mind, we present \dset, a dataset of \textbf{Multi}lingual \textbf{V}ideos of \textbf{E}vents with aligned \textbf{N}atural \textbf{T}ext that contains 2,396 diverse, event-centric videos and text descriptions that reflect the distribution of news content online. The videos are grounded in natural language video descriptions and long-form text documents, and the data spans 260 current events across over forty countries. The content in \dsetsp is collected in five target languages: Arabic, Chinese, English, Korean, and Russian, and as the multilinguality is organic, the data is less likely to suffer from translation bias. We provide an illustration of the dataset's contents in Figure \ref{fig:title}: Each natural language query (describing a video of a current event) is paired with grounding text documents and a unique corresponding video. We use \dsetsp to explore and characterize the variety of event-centric videos available online and illustrate the importance of leveraging these different video types when building multimodal information systems.

Citizen journalism, the most notable example being Wikipedia \cite{glott2010wikipedia}, has emerged alongside other online news sources as a method for curating comprehensive summaries of events. Work in natural language processing has considered the problem of automating this process by training models to generate informative reports using online source materials \cite{lewis2021paq, sciavolino2021simple, qian2023webbrain}. We use \dsetsp to explore how this process can be extended to incorporate multimodal sources of evidence. As a first step in this direction, we consider the task of video retrieval on \dset, through which a model learns to retrieve multimodal source material given a natural language event description. This task differs from prior video retrieval benchmarks \cite{chen2011collecting, xu2016msr, anne2017localizing, miech2019howto100m, wu2023large} as the videos in \dsetsp vary widely in length and content presentation, are multilingual, and can involve significant amounts of on-screen text. In addition to multilingual natural language captions for each video, we provide full text documents that ground the events and serve as more complex retrieval queries.

In summary, our contributions are:

\begin{enumerate}
\item We present \dset, a multimodal, multilingual information retrieval dataset of grounded videos depicting current events. The dataset targets five languages and covers a range of online video formats beyond traditional news broadcasts.

\item Using \dset, we quantitatively illustrate the information presented by news videos and the differences in content between video formats, and qualitatively evaluate how multimodal coverage of an event can evolve over time.

\item We present MultiCLIP, a model for multilingual, event-centric video retrieval that serves as a baseline for video retrieval approaches on the task.
\end{enumerate}

\section{Related Work}\label{related_work}

\subsection{Video retrieval datasets}

Early video datasets generally contained short clips spanning narrow ranges of topics, such as the Microsoft Research Video Description Corpus \cite{chen2011collecting}. Video datasets spanning larger domains include MSR-VTT \cite{xu2016msr} and DiDeMo \cite{anne2017localizing}, although the lengths of these videos were still relatively short. The V3C dataset \cite{rossetto2019v3c, awad2020trecvid} offered longer video lengths while still spanning a wide range of topics such as news reports. A shift towards massive video datasets was instigated by HowTo100M \cite{miech2019howto100m}, which included over 130 million video clips belonging to one million narrated instructional videos. VaTeX \cite{wang2019vatex}, released in the same year, considered video retrieval from a multilingual context using caption translation. Additional multilingual video retrieval datasets include Rudder \cite{dabral2021rudder}, consisting of instructional videos for making toys with multilingual captions, MTVR \cite{lei2021mtvr}, which extended the TVR dataset \cite{lei2020tvr} by adding Chinese subtitles and queries, and Multi-HowTo100M \cite{huang2021multilingual}, which extended HowTo100M  by scraping YouTube for subtitles in up to 9 other languages. Recently, Chen et al. \cite{chen2023chinaopen} released the ChinaOpen dataset which contains a wide range of video-caption pairs originally produced in Chinese. Recent work has also considered the problem of interpreting text-heavy video content: Wu et al. \cite{wu2023large} and Jahagirdar et al. \cite{jahagirdar2023watching} introduced datasets that focus on within-video text and OCR annotations, including news broadcasts.

\subsection{Video retrieval methods}

The size of early video datasets allowed retrieval systems to rely on pre-extracted features from expert systems like action recognition models. As massive video datasets gained prominence, the video retrieval paradigm moved towards ad-hoc video-text feature extraction using large pretrained models. Dosovitskiy et al. \cite{dosovitskiy2020image} proposed using stand-alone transformer architectures for video understanding, and Bertasius et al. \cite{bertasius2021space} showed that applying space- and time-based self-attention independently improved performance. Bain et al. applied findings directly to video retrieval, training and evaluating transformer architectures on WebVid-2M \cite{bain2021frozen}. Radford et al. \cite{radford2021learning} introduced CLIP and showed that pretraining models to match captions to images can result in scalable models, and CLIP's applicability to video retrieval was demonstrated by Fang et al. \cite{fang2021clip2video} through their CLIP2Video model. More fine-grained modifications to CLIP were proposed. Wang et al. \cite{wang2022object} introduced "Object-aware Transformers", which extended video-text transformers to incorporate object-level annotations within video footage, and Ge et al. \cite{ge2022bridging} modified the pretraining task to involve teaching a vision-text model to answer multiple choice questions about a video. Bain et al. \cite{bain2022clip} adapted large image-text models to the task of long video retrieval by incorporating the weighted-mean of frame embeddings, and Wu et al. \cite{wu2023large} incorporated independent optical character recognition and embeddings into the encoder pipeline to explicitly model in-video text.

\subsection{Report generation using online sources}

A wide range of research has used online corpora for report generation tasks, including QA-pair and knowledge graph generation \cite{rajpurkar2016squad, rajpurkar2018know, yang2018hotpotqa, mou2021narrative, kwiatkowski2019natural, petroni2020kilt}. Notably, Lewis et al. \cite{lewis2021paq} introduced a method for automatically extracting question-answer pairs from large corpora of text documents, and applied this method to Wikipedia to produce the PAQ dataset. Some PAQ extensions have been multilingual --- Pisare et al. \cite{pisarevskaya2022wikiomnia} built the WikiOmnia QA dataset on Russian Wikipedia documents, and Rybak et al. \cite{rybak2023maupqa} produced a question-Wikipedia passage dataset in Polish. Recently, Qian et al. \cite{qian2023webbrain} extended the ideas in PAQ to construct WebBrain, a task in which a model must generate factual articles with references given a natural language query. In the multimodal domain, Reddy et al. and Chen et al. have considered the problem of open-domain QA for image-text data \cite{reddy2022mumuqa, chen2022murag}, with Chen et al. using Wikipedia to generate a multimodal dataset. In a similar vein, Li et al. propose a dataset for information extraction from multimedia articles \cite{li2020cross} and an extraction approach that can be used with text, image, and video content \cite{li2020gaia}.
\section{Dataset}\label{sec:dataset}

In this section we outline the \dsetsp collection process. The dataset includes 2,396 videos and corresponding text descriptions covering 260 current events grounded in 468 text documents, and includes content in Arabic, Chinese, English, Korean, and Russian. We first identify 260 visually salient current events spanning from 2013 to 2023, and assign a target language to each event. Then, for each event, we collect grounding text documents and a set of videos in the event's target language.

\subsection{Current event curation}

We consider four primary event categories for \dset: Disasters, political events, social events, and technology events. We include thirteen current events per category for each target language. We use Google Trends statistics to select these events, based on its tracking of term popularity based on internet activity by country. We construct lists of the top five countries with the most speakers of each target language and review the top trending topics on Google in each of these countries over the last ten years. We record topics and search phrases that corresponded to current events that (1) align with one of the predefined event categories and (2) have sufficient online video coverage. For categories that did not amass a sufficient list of current events per language through this process, we consult Wikipedia's yearly summaries of events to fill the remaining slots. Detailed statistics characterizing this set of current events are shown in Figure \ref{fig:data-overview}. As shown, the majority of selected events take place in the last few years, with only three taking place before 2016.

Also shown in Figure \ref{fig:data-overview}, there is not a bijective mapping between the language used in event coverage and the country the event took place in. The language and country are often related, e.g., Russian news content in MultiVENT predominantly takes place in Russia, but this is not true of all events in the dataset. For example, we include data in Chinese pertaining to the 2023 ATP tennis circuit in Dallas, Texas: At this event, tennis player Wu Yibing became the highest-ranked Chinese player in the history of the ATP rankings, and so the event received substantial Chinese news coverage. In cases such as this, news in multiple languages will heavily focus on the same current event, such as sports events and international political relations. We do not include the same event in multiple languages in MultiVENT by design, in contrast with data collection procedures used for efforts such as AIDA \cite{tracey2022study} which aim to cover a small collection of current events in many languages. 

Every current event in the dataset is grounded in an English natural language document and, if the event is tagged with a non-English language, an additional natural language document in that target language. First, we check if a full English Wikipedia article exists for the current event. If not, we manually find a Wikipedia article that includes a passage describing the event. If Wikipedia does not have a passage that appropriately grounds the event, then a news article in English is selected as a grounding document instead. This process is then repeated for the target language. The dataset includes 468 grounding articles in total: 313 are full Wikipedia articles, 104 are Wikipedia passages, and 51 are external articles.

\subsection{Video collection}

\begin{figure}
 \includegraphics[width=\textwidth]{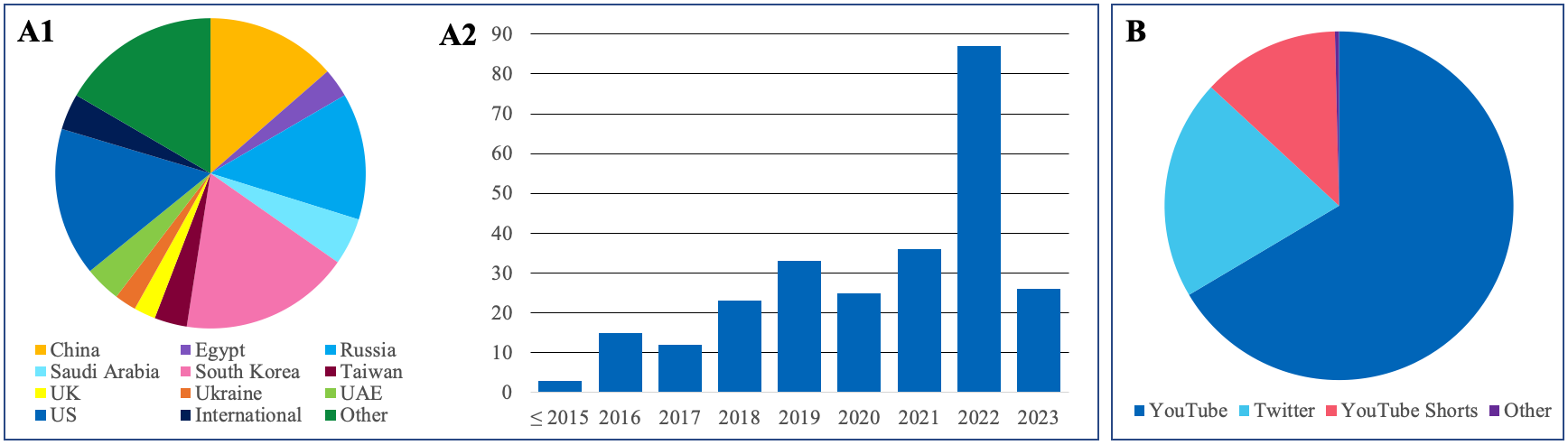}
  \caption{\textbf{A}: Statistics illustrating the distribution of current events selected for the dataset. (\textbf{A1}) depicts the general breakdown of countries in which each current event takes place. Many countries had a single current event, particularly small countries in the middle east and southeast Asia, and are consolidated into the "other" category for easier graph interpretation. (\textbf{A2}) shows the distribution of years during which the current events take place. \textbf{B}: Breakdown of data sources for the videos in the dataset. The majority of videos came from YouTube as it has a larger international audience and has existed longer than YouTube Shorts.}
 \label{fig:data-overview}
\end{figure}

We aim to collect visually and semantically distinct videos for each current event with an even split between firsthand witness accounts (e.g., first-person smartphone videos), amateur edited videos (e.g., vlogs), and professional news reports and compilations. Information regarding the resultant distribution of these categories and their semantic differences is included in Section \ref{sec:analysis-domain}. For each current event, we collect ten videos in the current event's target language. We search YouTube and Twitter for these videos using target keywords collected from the Google Trends search and Wikipedia. After collecting the videos, we manually identify and remove duplicates, resulting in 2,396 videos in total. We do not include repeat videos, but sometimes professional news reports include firsthand footage that is already included as unedited footage in the dataset. In these cases, we keep both the news report and the original footage as the context and text metadata between the two are distinct. If the video has a natural language description, we tag the video with this description. If it does not, we use the video title as the tagged natural language description. We report the distribution of videos by source in Figure \ref{fig:data-overview}.

\section{Data analysis}\label{sec:analysis}

We present an analysis of \dsetsp to help characterize how online multimodal content contributes to our understanding of current events. We explore multimodal event coverage from three angles: (1) what kinds of information news videos contribute, (2) the differences in informative content provided by different types of news videos, and (3) how multimodal coverage of an event can evolve over time.

\subsection{Semantic information in video}

Visual data can provide rich, semantically nuanced details of an event that are not captured in text documents due to reporting bias, limitations of text, and document length limits. To characterize the complexity of these videos and the information they provide, we annotate a set of two hundred videos of disasters in \dsetsp to identify visual entities in the videos that help answer common "who, what, where"-type questions about the events they depict.

We present videos of disaster footage to local annotators and provide them with a set of event-centric questions derived from FrameNet's "disaster scenario" template \cite{ruppenhofer2016framenet}. We modify this template, designed to annotate the event semantics of text documents, to better cover the range of information provided by visual content. We instruct annotators to identify every on-screen entity (such as people, scrolling news headline banners, etc.) that might help answer one of these event-centric questions.

The template divides salient entities into six categories: The disaster itself ("what"), the location of the disaster ("where"), the time the disaster takes place ("when"), people affected by the disaster ("who") and first responders for the disaster, e.g., firefighters (also "who"), and any visible outcomes of the disaster. Not every category applies to both visual content and text: We exclude "where" and "when" from the set of categories that visual content should be annotated for (because identifiable depictions of "where" are present in almost every frame, and "when" in virtually none) and disaster outcomes from the set of text annotation categories, as textual examples of this category tend to involve full clauses, which complicate the annotation process.

We present the number of event-relevant entities that appear on-screen in these annotated videos in Table \ref{tab:annotations}. For each annotated entity, we additionally ask annotators to rate their certainty that the entity is directly related to the event described by the video's natural language description from 0\% to 100\%. We record these certainty scores in 20\% intervals, i.e. as 20\%, 40\%, 60\%, 80\%, or 100\%. The averages of the linguists' confidence rankings by entity type are listed in Table \ref{tab:uncertainty}.

\begin{table*}[t]
 \centering
   \caption{Mean number of visual entities and in-scene text references (written text displayed within a video) present per video in a subset of 210 disaster videos from the current events dataset. We omit "where" and "when" entities from the visual content counts as "where" visual content technically appears in every frame and there are few types of visual evidence for "when" questions. We omit "outcomes" from the text references as an outcome by itself is a full event that is difficult to localize in text (this field is omitted from the FrameNet event template analogue for text documents).}
   \vspace{2mm}
  \begin{tabular}{lcccc}
  \thickhline
   & \textbf{Visual entities} & \textbf{Text references} & \textbf{Total} \\
  \toprule
     {Disaster ("What")} & 1.25 & 1.37 & \textbf{2.62} \\
     {Place of occurrence ("Where")} & - & 1.54 & \textbf{1.54} \\
     {Time of occurrence ("When")} & - & 0.77 & \textbf{0.77} \\
     {Affected people ("Who")} & 1.22 & 0.54 & \textbf{1.76} \\
     {First responders ("Who")} & 1.13 & 0.50 & \textbf{1.63} \\
     {Disaster outcomes} & 1.00 & - & \textbf{1.00} \\
     \toprule
     {Total} & \textbf{4.60} & \textbf{4.72} & \textbf{9.32} \\
    \thickhline
  \end{tabular}
  \label{tab:annotations}
\end{table*} 

\begin{table*}[t]
 \centering
   \caption{Mean annotator certainty scores partitioned on entity type based on the annotations used for Table \ref{tab:annotations}. 0.20 certainty indicates that the annotator is 20\% sure that the annotated entity helps answer the tagged question about the described event, while 1.00 certainty indicates that the annotator is completely sure that the entity helps answer the tagged question about the event.}
   \vspace{2mm}
  \begin{tabular}{lccccccc}
  \thickhline
   & \textbf{Disaster} & \textbf{Where} & \textbf{When} & \textbf{AP} & \textbf{FR} & \textbf{Outcomes} & \textbf{All} \\
  \toprule
     {Visual content} & .787 & - & - & .716 & .765 & .798 & .830 \\
     {Text content} & .931 & .907 & .929 & .856 & .836 & - & .900 \\
     \toprule
     {Average} & .862 & .907 & .929 & .759 & .787 & .798 & .865 \\
    \thickhline
  \end{tabular}
  \label{tab:uncertainty}
\end{table*}

As shown in Table \ref{tab:annotations}, each video contains an average of 9.32 informative visual entities that pertain to the event in question. About half of these entities are purely visual, and half are within-video text that can be identified with an optical character recognition model. As indicated by Table \ref{tab:uncertainty}, purely visual entities are more ambiguous than the text content shown onscreen alongside it, which aligns with past research that explores the difficulty humans have in interpreting visual content depicting complex events \cite{sandersambiguous}.

\subsection{Video content by domain}\label{sec:analysis-domain}

As described in Section \ref{sec:dataset}, we collect three main types of videos: Official news broadcasts, edited video footage, and raw, unedited footage. Of the 210 videos in the annotation set reported in Table \ref{tab:annotations}, 53\% are news broadcasts, 11\% are edited footage, and 36\% are raw footage. To quantify the difference in information presented by these different video types, we take the video annotations shown in Table \ref{tab:annotations} and partition these annotations by video type. We present the results in Table~\ref{tab:partitioned-anns}.

\begin{table*}[t]
 \centering
   \caption{Mean number of visual entities and in-scene text references present per video, partitioned on video type. Same 210 video subset is used for analysis as that used for the analysis shown in Table~\ref{tab:annotations}.}
   \vspace{2mm}
  \begin{tabular}{lccccccccc}
  \thickhline
   & \multicolumn{3}{c}{\textbf{News coverage}} & \multicolumn{3}{c}{\textbf{Edited footage}} & \multicolumn{3}{c}{\textbf{Raw footage}} \\
  \toprule
   & {Vis.} & {Text} & {Total} & {Vis.} & {Text} & {Total} &{Vis.} & {Text} &{Total} \\
  \toprule
     {Disaster ("What")} & 1.42 & 2.38 & \textbf{3.80} & 1.14 & 0.41 & \textbf{1.55} & 1.05 & 0.17 & \textbf{1.22} \\
     {Place ("Where")} & - & 2.51 & \textbf{2.51} & - & 0.59 & \textbf{0.59} & - & 0.39 & \textbf{0.39} \\
     {Time ("When")} & - & 1.26 & \textbf{1.26} & - & 0.41 & \textbf{0.41} & - & 0.16 & \textbf{0.16} \\
     {Affected people ("Who")} & 1.48 & 0.94 & \textbf{2.42} & 1.18 & 0.23 & \textbf{1.41} & 0.86 & 0.04 & \textbf{0.90} \\
     {First responders ("Who")} & 1.73 & 0.78 & \textbf{2.51} & 1.36 & 0.45 & \textbf{1.81} & 0.17 & 0.12 & \textbf{0.29}  \\
     {Disaster outcomes} & 1.28 & - & \textbf{1.28} & 0.77 & 0.27 & \textbf{1.04} & 0.67 & - & \textbf{0.67} \\
     \toprule
     {Total} & \textbf{5.91} & \textbf{7.87} & \textbf{13.78} & \textbf{4.45} & \textbf{2.36} & \textbf{6.81} & \textbf{2.75} & \textbf{0.88} & \textbf{3.63} \\
    \thickhline
  \end{tabular}
  \label{tab:partitioned-anns}
\end{table*}

As shown by the results, news broadcasts depict the most relevant semantic information, followed by edited footage. This is particularly apparent when considering text content alone. On average, news coverage contains almost 9 times as much relevant on-screen text content than raw footage, and over three times more than edited footage. Visual content differences were less drastic, but news content still had two times more visual content than raw footage and 1.3 times more than edited footage. The difference in visual content between news coverage and edited footage is possibly due to average video length and the quality of the video curation --- oftentimes, unprofessionally edited footage only draws from one source whereas news coverage draws from many.

\subsection{Information evolution}

\begin{figure*}
 \includegraphics[width=\textwidth]{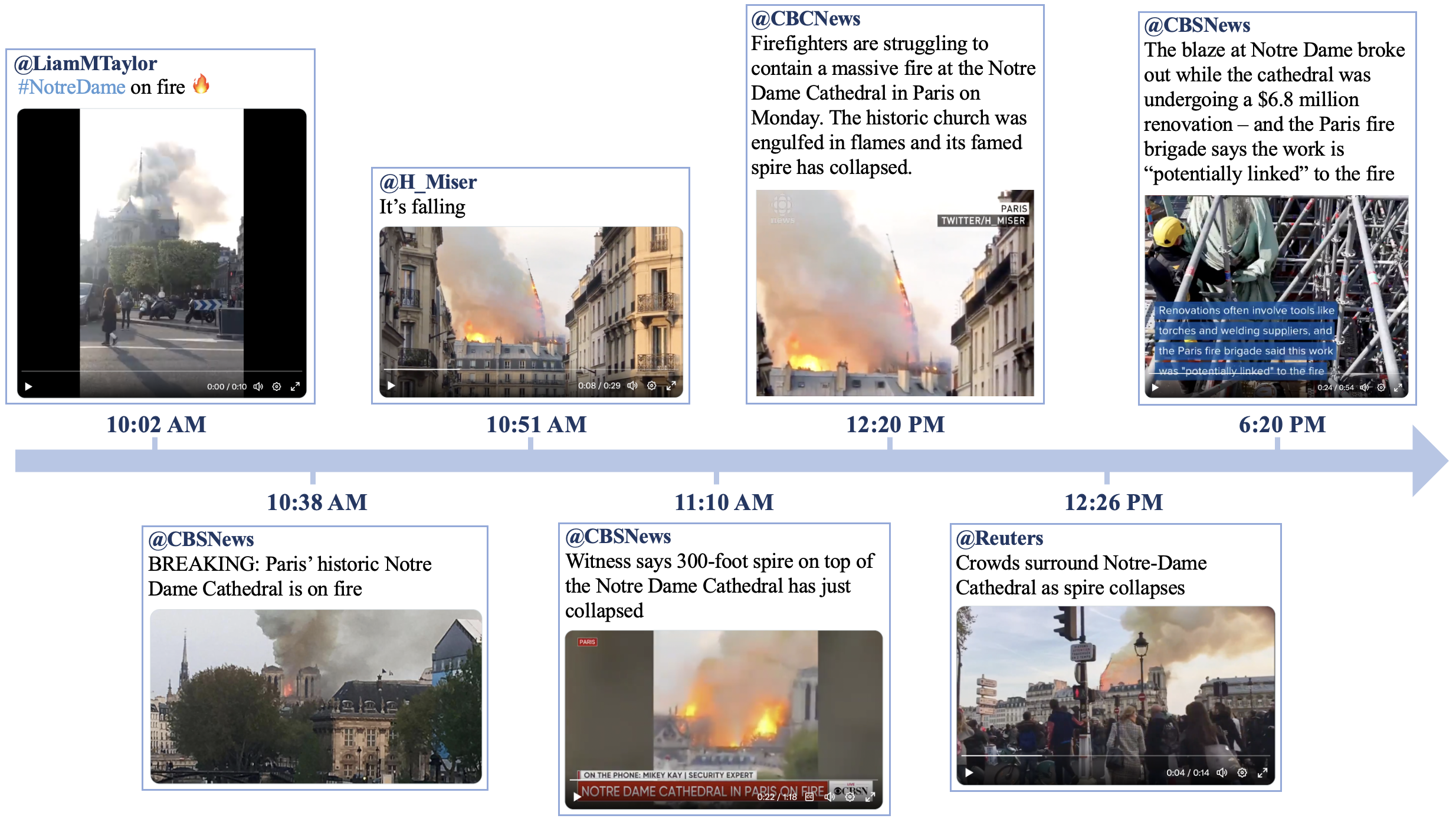}
  \caption{Snapshot of video news coverage of the 2019 Notre Dame fire news cycle (an event in \dset). The fire and the fallen spire were initially reported through first-person social media video uploads (at 10:02 AM and 10:51 AM, respectively) and then later broadcast by news organizations in more detail (10:38 AM, 11:10 AM, 12:20 PM, 12:26 PM). Some news coverage (12:20 PM) directly used first-person social media footage (10:51 AM). Hours later, news agencies uploaded more complete news stories with details and context (6:20 PM). This data suggests that it is important for models to learn from both first-person videos and official news coverage at various points in the news cycle to fully construct a factual model of the event, especially if the model is attempting to construct an event model while information develops online.}
 \label{fig:news-development}
\end{figure*}

As shown in Table \ref{tab:partitioned-anns}, first-person footage is often opaque compared to professional coverage. However, comprehensive coverage often builds on earlier, less informative coverage. This can be seen in news cycles for slowly unfolding events and for sudden, unexpected events that take time to assess. This is illustrated in Figure \ref{fig:news-development}, which shows a snapshot of the 2019 Notre Dame fire news cycle and demonstrates how unedited and poorly curated footage, often first-person witness accounts on social media, can be instrumental in the construction of our collective understanding of events. So, we propose that teaching models to understand different video formats, despite clear discrepancies in the amount of information they present, is important for developing robust systems.







\section{Experiments}\label{experiments}


\subsection{Approach}
\label{approach}

We consider the problem of teaching a model to map multilingual, natural language queries to multilingual video clips. Specifically, we consider a video set $V$ and query set $T$ with a indicator mapping function $f$ that returns whether a query $t\in T$ describes a video $v\in V$. The model $h$ is provided with the full set of videos $V$ and a text query $t\in T$, and for each video $v\in V$ returns the probability that $t$ describes $v$, or $h(v,t)=\mathbb{P}[f(v,t)=1]$. When there is a bijective mapping between queries and videos  (e.g., when using video descriptions as queries), the model is evaluated on its recall when considering the top 1, 5, and 10 ranked videos (R@1, R@5, and R@10), as well as the median rank (MedR). When a given query may describe multiple videos, (e.g., when using event descriptions as queries), we instead evaluate the model on its precision given the top 1, 5, and 10 ranked videos (P@1, P@5, and P@10). We define these metrics as:

$$\text{Given } \hspace{1mm} S:= \argmax_{V' \subseteq V : |V'| = k}\hspace{1mm}\sum_{v\in V'}h(v,t) \text{,}$$

$$\text{R@}k=\frac{|\{s\in S \hspace{1mm}:\hspace{1mm}f(s,t)=1\}|}{|\{v\in V \hspace{1mm}:\hspace{1mm}f(v,t)=1\}|}\hspace{3mm} \text{and } \hspace{3mm} \text{P@}k=\frac{|\{s\in S \hspace{1mm}:\hspace{1mm}f(s,t)=1\}|}{k}\text{.}\hspace{3mm}$$

\subsection{Model architecture and training}

We introduce MultiCLIP, a multilingual baseline for video retrieval on \dset. We base our architecture on the pretrained LAION CLIP ViT-H/14 frozen XLM-Roberta-Large model \cite{cherti2022reproducible}, which jointly trains an image and text encoder on text-image data to learn to pair images with their captions. At test time, it produces a zero-shot linear layer based on the test input's visual features through which natural language captions can be passed in. The model architecture contains a vision encoder based on a ViT architecture \cite{dosovitskiy2020image} and a text encoder based on the the multilingual XLM Roberta large model \cite{conneau2019unsupervised}. A full overview of the CLIP architecture and pretraining can be found in the original paper \cite{radford2021learning}. 

In experiments using MultiCLIP, we first tokenize text descriptions using the XLM-Roberta-Large tokenizer, containing a vocabulary of over 250,000 words, and pass the tokens into MultiCLIP which produces a text embedding of size 1024. Next, we uniformly sample videos at a rate of 12 frames per video with an input size of 224x224, which the model uses to create a frame embedding of size 1024. To incorporate multilinguality into the model's frame-level features, we use a ViT architecture trained with a contrastive objective over multilingual image-caption pairs from the LAION-5B dataset \cite{schuhmann2022laion}, which is constructed from the Common Crawl archive using images and their alt-text to produce a multilingual image-text dataset with over 100 languages. We mean pool the frame embeddings to produce a final video embedding, and use the text and video features to compute a similarity matrix of videos and descriptions.

\subsection{Retrieval baselines}
We first evaluate MultiCLIP on the existing video retrieval task MSR-VTT \cite{xu2016msr} using the recall metrics described in Sec. \ref{approach} alongside contemporary video retrieval approaches (FrozenInTime \cite{bain2021frozen}, Clip2Video \cite{fang2021clip2video}, InternVideo \cite{wang2022internvideo}, and MPLUG-2 \cite{xu2023mplug}). Results on MSR-VTT's validation set are reported in Table \ref{tab:0}. The results indicate  MultiCLIP  performs well on standard video retrieval tasks, matching performance of separate text/vision pipeline models released within the last two years. It performs better than existing models that use separate text and vision pipelines (FrozenInTime \cite{bain2021frozen} and Clip2Video \cite{fang2021clip2video}), but not as well as models that use larger architectures involving multimodal encodings (InternVideo \cite{wang2022internvideo} and MPLUG-2 \cite{xu2023mplug}).

\begin{table*}[t]
 \centering
   \caption{MultiCLIP evaluated alongside existing video retrieval approaches on the video retrieval benchmark MSR-VTT. Results indicate that MultiCLIP performs adequately on existing retrieval tasks, achieving comparable results to existing models. It does not perform as well as architectures that use multimodal transformers for joint encodings such as InternVideo and MPLUG-2.\\}
  \begin{tabular}{lccccc}
  \thickhline

   \textbf{Method} & \textbf{Year} & \textbf{Rank@1} & \textbf{Rank@5} & \textbf{Rank@10}  \\
  \hline
    FrozenInTime \cite{bain2021frozen} & 2021 & 32.5 & 61.5 & 71.2 \\
    Clip2Video \cite{fang2021clip2video} & 2021 & 29.8 & 55.5 & 66.2 \\
    InternVideo \cite{wang2022internvideo} & 2022 & \textbf{55.2} & \textbf{79.6} & \textbf{87.5}   \\
    MPLUG-2 \cite{xu2023mplug} & 2023 & 53.1 & 77.6 & 84.7 \\
    \textbf{MultiCLIP} & 2023 & 38.4 & 70.1 & 82.7  \\
    \thickhline
  \end{tabular}
  \label{tab:0}
\end{table*}

\subsection{\dsetsp retrieval}
We now evaluate MultiCLIP and related retrieval approaches on \dset. We first use multilingual video descriptions as queries, and then we use English event summaries taken from the grounding text documents, meaning that one text query maps to up to ten videos. The event queries are selected by taking one to two sentences from each English event text document that describes the event most holistically. We exclusively use English queries for this section, as our annotators fluent in the other languages were not available for this task. In addition to MultiCLIP, we consider a set of contemporary video retrieval models with lightweight architectures (FrozenInTime \cite{bain2021frozen}, CLIP2Video \cite{fang2021clip2video}, 
InternVideo \cite{wang2022internvideo}, and a pooled CLIP model using the same setup as MultiCLIP without the additional multilingual pretraining). We argue that lightweight architectures are most appropriate for evaluating a full, pairwise set of similarity scores between text and video data of large multimodal corpora. Results are reported, partitioned on language, in Table \ref{tab:1}.

We report the standard recall @ rank $k$ metric for retrieval on individual video queries, and precision @ rank $k$ for retrieval on event description queries. The results suggest that some existing video retrieval models may particularly struggle on this task, regardless of language. We hypothesize that this is due to a combination of the videos' length, complex semantic content, ambiguity, and frequent OCR content, as well as the long and often noisy video description queries.

While \dset as a whole poses challenges to existing models, it is also clear that multilingual data may significantly impact performance on models trained primarily on English content - all models suffer a performance loss when evaluated on multilingual content (even when using English queries, as shown by the event description query results). While MultiCLIP suffers a performance loss on this data as well, comparing the standard pooled CLIP model against MultiCLIP shows that training on multilingual data does mitigate this multilingual performance loss: The two models perform comparably on English data, but MultiCLIP performs better on the multilingual content, especially when multilingual queries are used.

\begin{table*}[t]
 \centering
   \caption{Results showing the retrieval performance of video retrieval methods alongside MultiCLIP on \dset. We use video descriptions and event descriptions as queries and partition results based on language. As shown, \dsetsp can be a difficult retrieval benchmark for video retrieval models even when considering only English, but the benefit of training on multilingual data is apparent when comparing MultiCLIP against the regular pooled CLIP model on non-English data.\\}
  \begin{tabular}{lccccccc}
  \thickhline
  & \multicolumn{4}{c}{Video description} & \multicolumn{3}{c}{Event description} \\
  \hline
   \textbf{Method}  & \textbf{R@1} & \textbf{R@5} & \textbf{R@10} & \textbf{MedR} & \textbf{P@1} & \textbf{P@5} & \textbf{P@10}\\
  \hline
  \multicolumn{8}{c}{English} \\ 
  \hline
    FrozenInTime \cite{bain2021frozen} & 6.5 & 20.0 & 28.4 & 53.0 & 42.3 & 34.6 & 26.9  \\
    {CLIP2Video} \cite{fang2021clip2video}& 41.3 & 71.8 & 80.4 & 2.0 & 96.2 & \textbf{96.9} & 73.3  \\
    {InternVideo} \cite{wang2022internvideo} & 53.8 & 83.1 & 88.7 & \textbf{1.0} & 94.2 & 93.5 & 79.6  \\
    CLIP (pooled) \cite{radford2021learning}  & \textbf{55.9} & 83.9 & 91.3 & \textbf{1.0} & 98.1 & 94.6 & \textbf{80.6}  \\
    \textbf{MultiCLIP}  & \textbf{55.9} & \textbf{84.5} & \textbf{92.3} & \textbf{1.0} & \textbf{100.0} & \textbf{96.9} & \textbf{80.6}  \\
    \hline 
  \multicolumn{8}{c}{Arabic + Chinese + Korean + Russian} \\ 
  \hline
    FrozenInTime \cite{bain2021frozen} & 0.5 & 1.2 & 2.5 & 793.5 & 29.8 & 22.6 & 17.6  \\
    {CLIP2Video} \cite{fang2021clip2video} & 2.4 & 7.2 & 10.5 & 166.5 & 14.4 & 9.4 & 7.3  \\
    {InternVideo} \cite{wang2022internvideo} & 5.7 & 13.9 & 19.8 & 91.0 & 79.3 & 71.0 & 55.7  \\
    CLIP (pooled) \cite{radford2021learning} &  6.2 & 15.9 & 22.4 & 79.5 & 83.7 & 73.3 & 58.2  \\
    \textbf{MultiCLIP} &  \textbf{32.6} & \textbf{64.7} & \textbf{79.5} & \textbf{3.0} & \textbf{85.6} & \textbf{76.4} & \textbf{61.5}  \\
    \thickhline
  \end{tabular}
  \label{tab:1}
\end{table*}

\section{Conclusion}\label{conclusion}

We introduce \dset, a multimodal, multilingual dataset grounded in natural language documents for event-centric video retrieval and information acquisition. This dataset consists of 2,396 videos covering 260 current events reported in five target languages (Arabic, Chinese, English, Korean, and Russian) paired with multilingual natural language video descriptions and long-form event-centric text documents. We use this dataset to characterize online news coverage and how models can use this online content for information acquisition. We propose a multilingual video retrieval benchmark using \dsetsp and present MultiCLIP, multilingual video retrieval model to serve as a baseline for the task. We evaluate this model and related retrieval approaches on MSR-VTT and \dsetsp to illustrate the importance of pretraining on multilingual data for evaluation on \dset. In future work, we aim to explore the effect that joint vision-OCR embeddings can have on video retrieval in text-heavy contexts. Also in future work, a RePAQ-adjacent system \cite{lewis2021paq} for automatically extracting question-answer pairs from video content and video-document pairs could be developed and applied to \dset. Through this, a framework for teaching models to perform open-domain question-answering tasks with multimodal background corpora could be established, expanding the domain of questions a model can answer.


{\small
\bibliographystyle{plain}
\bibliography{references}
}

\newpage
\appendix
\end{document}